\documentclass[fl eqn,twoside]{article}
\date{}
\usepackage{amsmath}
\usepackage{amsfonts}
\usepackage{amssymb}
\usepackage{graphicx}
\date{\today}
\begin{document}

\title{{\bf{Path integral action in the generalized uncertainty principle framework}}}

\author{
{\bf {\normalsize Sukanta Bhattacharyya}$^{a}
$\thanks{sukanta706@gmail.com}},
{\bf {\normalsize Sunandan Gangopadhyay}
$^{b}$\thanks{ sunandan.gangopadhyay@bose.res.in, sunandan.gangopadhyay@gmail.com}}\\
$^{a}$ {\normalsize Department of Physics, West Bengal State University, Barasat, Kolkata 700126, India}\\
$^{b}$ {\normalsize Department of Theoretical Sciences}\\
{\normalsize S.N. Bose National Centre for Basic Sciences}\\
{\normalsize JD Block, Sector III, Salt Lake, Kolkata 700106, India}\\
}
\date{}
\maketitle
\begin{abstract}
\noindent Various gedanken experiments of quantum gravity phenomenology in search of a complete theory of gravity near the Planck scale indicate a modification of the Heisenberg uncertainty principle to the generalized uncertainty principle (GUP). This modification leads to nontrivial contributions on the Hamiltonian of a nonrelativistic particle moving in an arbitrary potential. In this paper we study the path integral representation of a particle moving in an arbitrary potential using the most general form of the GUP containing both the linear and quadratic contributions in momentum. First we work out the action of the particle in an arbitrary potential and hence find an upper bound to the velocity of a free particle. This upper bound interestingly imposes restrictions on the relation between the GUP parameters $\alpha$ and $\beta$. Analysis shows that $ \beta > 4 \alpha^2$. We then deduce the mathematical expressions of classical action and the quantum fluctuations for both free particle and the harmonic oscillator systems.   
\end{abstract}
\section{Introduction}
\noindent General relativity (GR), proposed by Einstein in 1915 \cite{en} can explain and predict a large number of physical phenomena in astrophysics and cosmology. On the other hand, quantum mechanics (QM) is the most successful candidate to describe the dynamics of a particle in the microscopic world. Since GR is a classical theory, it fails to characterize the universe near the Planck scale. Though these two fundamental theories (GR and QM) are very successful in their own-domain, a complete quantum theory of gravity to explore the universe close to the Planck epoch is absolutely essential. Interestingly substantial investigation of various theories of quantum gravity such as string theory (ST) \cite{da}, loop quantum gravity (LQG) \cite{cr,sc}, noncommutative (NC) geometry \cite{fg} and some gedanken experiments in quantum gravity phenomenology (QGP) hints at the existence of a minimal length, namely, the Planck-length. This fundamental hypothesis of the observer independent Planck-length together with the Heisenberg uncertainty principle (HUP), one of the main pillars of QM, leads to a modification of the HUP \cite{ca}. This modification is well known in the literature as the generalized uncertainty principle (GUP). A large area in theoretical physics which includes black hole thermodynamics \cite{rj}-\cite{sg1}, various quantum system like particle in a box, Landau levels, simple harmonic oscillator \cite{sd}-\cite{sd2}, have been extensively studied under the GUP framework. Recently studies have been done to investigate the path integral formalism of a nonrelativistic particle in presence of the GUP \cite{sp,sb}. In \cite{sb}, the Feynman propagator of a particle under any arbitrary potential has been calculated using the simplest form of the GUP, in which the modification to the HUP involves a term proportional to the quadratic in momentum. This modification of the HUP, proportional to the quadratic in momentum is motivated by black hole physics and string theory. However, theories of doubly special relativity (DSR) suggest that there can be a modification involving a term linear in momentum. Hence combining both possibilities the most general form of the GUP has been first introduced in \cite{sd2}. In this paper we want to explore the path integral formalism for a nonrelativistic particle moving under any arbitrary potential using this form of the GUP which contains both the linear and quadratic modifications in momentum.

\noindent The modified uncertainty principle between the position $q_i$ and its conjugate momentum $p_j$ incorporating both the contributions, linear and quadratic in momentum is given by \cite{sd},
    
\begin{equation}
\Delta q_i \Delta p_i\geq~ \frac{\hbar}{2}\left[1-\alpha \left\langle p+\frac{p_i p_i}{p}\right\rangle-(\alpha^2-\beta)\left((\Delta p)^2+{\langle p\rangle}^2\right)-(\alpha^2-2\beta)\left(({\Delta p_i})^2+{\langle p_i\rangle}^2\right)\right]
\label{gupvp}
\end{equation}
\noindent  where ${{p}}^2\equiv{\vert\vec{{p}}\vert}^2= \eta_{ij}p^{i}p^{j}$; $i,j=1,2,3$, the parameters $\alpha$ and $\beta$ bear the signature of the GUP, and are defined as $\alpha=\alpha_{0}/(M_{Pl}c)$ and $\beta=\beta_{0}/(M_{Pl}c)^2$, $M_{Pl}$ being the Planck mass and $c$ is the speed of light in free space. The dimensions of $\alpha$ and $\beta$ are $(momentum)^{-1}$ and $(momentum)^{-2}$ respectively. The above uncertainty principle is consistent with the following modified Heisenberg algebra
\begin{eqnarray}
[q_i,p_j]=i\hbar\left[\delta_{ij}-\alpha\left(\delta_{ij}p+\dfrac{p_i p_j}{p}\right)
+\beta\left(\delta_{ij}p^2+2p_i
p_j\right)-\alpha^2(\delta_{ij}p^2+p_i p_j)
\right]~.
\label{gupu}
\end{eqnarray}
\noindent It can be easily shown that the above commutator follows the Jacobi identity, $[q_i, q_j]=[p_i, p_j]=0~$. The modified variables $(q_j, p_j)$ can be expressed in terms of the usual variables $(q_{0j},p_{0j})$ in such a way that they obey the commutation relation (\ref{gupu}). Hence we have 
\begin{eqnarray}
q_{j} &=& q_{0j}~,\nonumber\\
p_{j} &=& p_{0j}(1-\alpha{p_0}+\beta{{p_0}}^2)~,
\label{GUPR}
\end{eqnarray}

\noindent where $(q_{0j}, p_{0j})$ satisfies the usual commutation relation $[q_{0i}, p_{0j}]=i \hbar \delta_{ij}$. With the relation (\ref{GUPR}) in hand, we now construct the GUP modified Hamiltonian describing a particle moving under any arbitrary potential $V(q)$. In the subsequent discussion we shall work in one spatial dimension. Therefore using eq.(\ref{GUPR}), the Hamiltonian of the particle in an arbitrary potential $V(q)$ upto order $\mathcal{O}(\alpha^2, \beta)$ reads

\begin{eqnarray} 
\hat H=\frac{ \hat p_{0}^{2}}{2m}-\frac{\alpha}{m} \hat p_{0}^{3}+\frac{1}{m}\left(\frac{\alpha^2}{2}+\beta\right)\hat p_{0}^{4}+V(\hat q) +\mathcal O( \alpha \beta, \alpha^3, \beta^2)~.
\label{GUPH}
\end{eqnarray}
\noindent 
\noindent We are now ready to construct the path integral representation of a particle moving in an arbitrary potential in the GUP modified Hamiltonian (\ref{GUPH}). Note that we have used the most general form of the GUP incorporating both the contributions linear and quadratic in momentum. In this paper first we calculate the classical action and the explicit form of the propagation kernel for a particle moving under any arbitrary potential using the Hamiltonian  (\ref{GUPH}) in section I. In this section, we derive the equation of motion of the particle which gives an interesting relation between the GUP parameters $\alpha$ and $\beta$. In the next section II, we evaluate the explicit form of the propagation kernel both for the free particle and the harmonic oscillator. Then we proceed to section III to calculate the quantum fluctuations of the propagation kernel for the harmonic oscillator. Finally, we summarize our results in the concluding section.

\section{Propagation kernel for an arbitrary potential}
\noindent To investigate the path integral formalism of a particle described by the GUP modified Hamiltonian (\ref{GUPH}), we write the general form of the propagation kernel as 


\begin{eqnarray}
\langle q_{f}, t_{f}| q_{0}, t_{0} \rangle= \lim_{n \to \infty} \int_{-\infty}^{+\infty} \prod_{j=1}^{n} dq_{j} \langle q_{f},t_{f}| q_{n}, t_{n} \rangle \langle  q_{n}, t_{n}|.~.~.|q_{1},t_{1} \rangle \langle     q_{1},t_{1}|q_{0},t_{0} \rangle~.
\label{propafree}
\end{eqnarray}
\noindent Then we compute the propagator over a small segment in the above path integral. Here we use the following completeness relation  

\begin{eqnarray}
\int_{-\infty}^{+\infty} dp~ |p\rangle \langle p|= \mathbf{1}~.
\label{cr}
\end{eqnarray}
\noindent Hence the Hamiltonian given by eq.(\ref{GUPH}) along with the above completeness relation (\ref{cr}) gives the infinitesimal propagator, which reads
\begin{eqnarray}
\langle q_{j+1},t_{j+1}|q_{j},t_{j}\rangle&=& \langle q_{j+1}|e^{-\frac{i}{\hbar}\hat H\tau}|q_{j}\rangle \nonumber\\
&=&\langle q_{j+1}|1-\frac{i}{\hbar}\hat H\tau+ \Theta(\tau^2)|q_{j} \rangle \nonumber\\&&
=\int_{-\infty}^{+\infty} \frac{dp_{j}}{2\pi\hbar}~~ e^{\frac{i}{\hbar} p_{j}(q_{j+1}-q_{j})} e^{-\frac{i}{\hbar}\tau \left(\frac{p_{j}^{2}}{2m}-\frac{\alpha}{m}p_{j}^{3}+\frac{1}{m}\left(\frac{\alpha^2}{2}+\beta\right)p_{j}^{4}+V(q_j)\right)}+\mathcal O(\tau^2)~.
\label{propafree2}
\end{eqnarray}
\noindent Substituting the above expression in eq.(\ref{propafree}), the propagation kernel takes the form (apart from a constant factor)
\begin{eqnarray}
\langle q_{f}, t_{f}| q_{0}, t_{0}\rangle =\lim_{n \to \infty} \int_{-\infty}^{+\infty} \prod_{j=1}^{n} dq_{j} \prod_{j=0}^{n} dp_{j} ~exp \left(\frac{i}{\hbar} \sum_{j=0}^{n} \left[p_{j}(q_{j+1}-q_{j})-\tau \left\{\frac{p_{j}^{2}}{2m}-\frac{\alpha}{m}p_{j}^{3}+\frac{1}{m} \left(\frac{\alpha^2}{2}+\beta\right)p_{j}^{4}+V(q_j)\right\}\right]\right).
\label{propa2}
\end{eqnarray}
Taking $\tau \to 0$ limit, the propagation kernel given by eq.(\ref{propafree}) in the path integral representation reads 
\begin{eqnarray}
\langle q_{f}, t_{f}| q_{0}, t_{0}\rangle=\int \mathcal{D}q ~\mathcal{D}p~~ exp \left(\frac{i}{\hbar} \mathcal{A}\right)
\label{psprop}
\end{eqnarray} 
where the phase-space action $\mathcal{A}$ is given by 
\begin{eqnarray}
\mathcal{A}= \int_{t_0}^{t_f} dt \left[p\dot q- \left\{\frac{p^{2}}{2m}-\frac{\alpha}{m}p^{3}+\frac{1}{m}\left(\frac{\alpha^2}{2}+\beta\right)p^{4}+V(q)\right\}\right]~, ~t_f-t_0=T~.
\label{psaction}
\end{eqnarray}
\noindent Now we compute the momentum integral in eq.(\ref{propafree2}). Evaluating this keeping terms upto $\mathcal{O}(\alpha^2, \beta)$ yields
\begin{eqnarray}
\langle q_{j+1},t_{j+1}~|~q_{j},t_{j}\rangle&=&\sqrt{\frac{m}{2\pi i\hbar\tau}}
\left[1+\left\{\frac{i \alpha m^2 (q_{j+1}-q_{j})^3}{\hbar \tau^2}+\frac{1}{2}\left(\frac{i \alpha m^2 (q_{j+1}-q_{j})^3}{\hbar \tau^2}\right)^2 \right\}+ \frac{3 \alpha m (q_{j+1}-q_{j})}{\tau} \right. \nonumber\\ && \left. + \frac{3 i \hbar m \beta}{\tau}-\frac{6 i \hbar m \alpha^2}{\tau} -6 m^2 \left(\frac{\alpha^2}{2}+\beta \right) \left(\frac{q_{j+1}-q_{j}}{\tau}\right)^2- \frac{i m^3 \beta(q_{j+1}-q_{j})^4}{\hbar \tau^3} m \right. \nonumber\\ && \left.  +\frac{7 i \alpha^2 m^3 (q_{j+1}-q_{j})^4}{\hbar \tau^3 }+ \frac{45 \alpha^2 m^2(q_{j+1}-q_{j})^2}{2 \tau^2}\right] \times exp\left({\frac{im(q_{j+1}-q_{j})^{2}}{2\hbar\tau}}+\frac{i}{\hbar}\tau V(q_j)\right)\nonumber\\
&=& \sqrt{\frac{m}{2\pi i\hbar\tau}}
\left[1- \frac{i}{\hbar} \left\{ \frac{ m^3 \beta (q_{j+1}-q_j)^4}{\tau^3} - \frac{4 \alpha^2 m^3 (q_{j+1}-q_j)^4}{\tau^3} \right\} +\frac{3 \alpha m (q_{j+1}-q_j)}{\tau}\right. \nonumber\\ && \left.  +\frac{3\beta i\hbar m}{\tau}-\frac{6 i \hbar m \alpha^2}{\tau} -6 m^2 \left(\frac{\alpha^2}{2}+ \beta \right) \left(\frac{q_{j+1}-q_j}{\tau}\right)^2 + \frac{45 \alpha^2 m^2}{2} \left( \frac{q_{j+1}-q_j}{\tau}\right)^2 \right]\nonumber\\&&
\times  exp\left({\frac{im(q_{j+1}-q_{j})^{2}}{2\hbar\tau}}+\frac{i}{\hbar}\tau V(q_j)-\frac{i \alpha m^2 (q_{j+1}-q_j)^3}{\hbar \tau^2}\right) \nonumber\\
&=&  \sqrt{\frac{m}{2\pi i\hbar\tau}} \left[\frac{3 \alpha m (q_{j+1}-q_j)}{\tau}+\frac{3\beta i\hbar m}{\tau}-\frac{6 i \hbar m \alpha^2}{\tau} -6 m^2 \beta \left(\frac{q_{j+1}-q_j}{\tau}\right)^2 \right. \nonumber\\ && \left.+ \frac{39 \alpha^2 m^2}{2} \left( \frac{q_{j+1}-q_j}{\tau}\right)^2 \right] \times  exp\left({\frac{im(q_{j+1}-q_{j})^{2}}{2\hbar\tau}}+\frac{i}{\hbar}\tau V(q_j)-\frac{i \alpha m^2 (q_{j+1}-q_j)^3}{\hbar \tau^2} \right.  \nonumber\\ && \left. -\frac{i \beta m^3 (q_{j+1}-q_j)^4}{\hbar \tau^3} + \frac{4 i \alpha^2 m^3 (q_{j+1}-q_j)^4}{\hbar \tau^3}\right)
+\mathcal{O}(\beta^2)
. \label{infpropafree3}
\end{eqnarray}
Using the above result in eq.(\ref{propa2}), we obtain the propagation kernel upto a constant factor as 
\begin{eqnarray}
\langle q_{f},t_{f} |q_{0}, t_{0} \rangle &=& \int_{-\infty}^{+\infty} \prod_{j=1}^{n} dq_{j}~~exp\left(\frac{i}{\hbar}\sum_{j=0}^{n} \tau \left[ \frac{m}{2}\left(\frac{q_{j+1}-q_{j}}{\tau}\right)^2\left\{1+2 \alpha m \left(\frac{q_{j+1}-q_j}{\tau}\right)+ 8 \alpha^2 m^2 \left(\frac{q_{j+1}-q_j}{\tau}\right)^2 \right. \right. \right. \nonumber\\ && \left. \left. \left. -2\beta m^2\left(\frac{q_{j+1}-q_{j}}{\tau}\right)^2\right\}-V(q_j)\right]\right)~.
\label{propagafree1}
\end{eqnarray}
\noindent To get the configuration space path integral representation of a particle moving in an arbitrary potential $ V(q)$, we take the limit $ \tau \to 0$. This gives  
\begin{eqnarray}
\langle q_{f},t_{f} |q_{0}, t_{0} \rangle =\tilde{F}(T, \alpha, \beta) \int \mathcal{D}q~~e^{\frac{i}{\hbar}S}~
\label{confipath}
\end{eqnarray}
\noindent where the action of the particle moving in the presence of an arbitrary  potential $V(q)$ in the configuration space is given by
\begin{eqnarray}
S=\int_{t_{0}}^{t_{f}} dt \left[\frac{m}{2} \dot q ^2\left(1+2 \alpha m \dot q + 8 \alpha^2 m^2 \dot q^2 - 2 \beta m^2 \dot q^2 \right)- V(q)\right]~.
\label{confiaction}
\end{eqnarray}
\noindent From the above action one can readily write down the Lagrangian to be
\begin{eqnarray}
L= \frac{m}{2} \dot q^2 \left( 1+ 2 \alpha m \dot q + 8 \alpha^2 m^2 \dot q^2- 2 \beta m^2 \dot q^2\right)- V(q)~.
\label{lagran}
\end{eqnarray}
\noindent Eqs.(\ref{confiaction}) and (\ref{lagran}) are the general forms of the action and the Lagrangian of a non relativistic particle moving under an arbitrary potential $V(q)$ in presence of the GUP. It is to be noted that here we take the generalized structure of the GUP (\ref{GUPR}) where both the linear and quadratic modifications in momentum $p_j$ are present. We now proceed to investigate the free particle and the harmonic oscillator systems.

\section{Free particle and HO system}
\noindent With the above results in hand, we now proceed to investigate the free particle and harmonic oscillator potential in this section.

\noindent For the free particle case we have $V(q)=0$. Hence from the action in eq.(\ref{confiaction}) one can easily find the classical equation of motion. This reads
\begin{eqnarray}
m\left(1+6 \alpha m \dot q + 48 \alpha^2 m^2 \dot q ^2 -12 \beta m^2 \dot q^2\right) \ddot q &=& 0  \nonumber\\  \implies \ddot q = 0 ~~~\textrm{or}~~~ \left(1+6 \alpha m \dot q + 48 \alpha^2 m^2 \dot q ^2-12 \beta m^2 \dot q^2\right) &=& 0~.
\label{eqm}
\label{classeqm}
\end{eqnarray}
\noindent Before going further we now analyze the above result. Interestingly both the possibilities indicate $\ddot q= 0$. Moreover the presence of the GUP puts a bound on the free particle velocity. The upper bound on the velocity of the free particle obtained from the action reads
\begin{eqnarray}
\dot q_{max}=\frac{-\alpha - \sqrt{(2\beta- 7 \alpha^2)}}{2m(4 \alpha^2-\beta)}~.
\label{freevel}
\end{eqnarray}
\noindent Note that in the limit $\alpha \to 0$, this maximum particle speed agrees with the result obtained in \cite{sb}. Now the free particle velocity given by eq.(\ref{freevel}) cannot be imaginary and must be finite. This restriction gives a relation between the GUP parameters $\alpha$ and $\beta$ which is

\begin{eqnarray}
\beta > 3.5 \alpha^2~~~~\textrm{and}~~~~ \beta \ne 4 \alpha^2~. 
\label{rel}
\end{eqnarray}
\noindent It should also be noted that we have taken the negative sign before the square root. The reason behind this choice is that there is no value of $\beta > 3.5 \alpha^2$ for which $\dot q_{max}$ is positive. With the negative sign before the square root, the positivity of $\dot q_{max}$ implies $\beta > 4 \alpha^2 $.
\noindent We now want to point out some relevant and interesting facts about the above relations. In \cite{sd}, the authors have showed that $\beta = 2 \alpha^2 $ (see Appendix I). But from our analysis we find that the relation $\beta = 2 \alpha^2 $ is not possible. From the analysis in \cite{sd} (see Appendix I), we have 
\begin{eqnarray}
\beta=(n+1) \alpha^2~.
\label{ab}
\end{eqnarray}
\noindent Therefore, eq.(\ref{rel}) together with eq.(\ref{ab}) gives the relation between $\alpha$ and $\beta$. This is an important result in our paper. 

\noindent We now calculate the classical action for the free particle in presence of the GUP. To do this first we have to solve $\ddot q=0$ imposing the boundary conditions that at $t=t_0$, $q=q_0$; $t=t_f$, $q=q_f$. The classical trajectory of the free particle then comes out to be
\begin{eqnarray}
q_{c}(t)=q_0 + \frac{q_f-q_0}{T}t~,~~~~~~ t_f-t_0=T~.
\label{clstra}
\end{eqnarray}
\noindent Substituting eq.(\ref{clstra}), in eq.(\ref{confiaction}) the classical action for the free particle in the presence of the GUP takes the form   
\begin{eqnarray}
S_{c}= \frac{m}{2 T} (q_{f}-q_{0})^2 \left[ 1+ 2 \alpha m \left(\frac{q_f-q_i}{T}\right)+ 8 \alpha^2 m^2 \left(\frac{q_f-q_i}{T}\right)^2 - 2 \beta m^2 \left(\frac{q_{f}-q_{0}}{T}\right)^2\right]~.
\label{clssaction}
\end{eqnarray}
\noindent Using the above expression for the classical action in eq.(\ref{confipath}), we obtain 
\begin{eqnarray}
\langle q_{f},t_{f} |q_{0}, t_{0} \rangle = \tilde{F}(T,\alpha, \beta)~e^{\frac{m}{2 T} (q_{f}-q_{0})^2 \left[ 1+ 2 \alpha m \left(\frac{q_f-q_i}{T}\right)+ 8 \alpha^2 m^2 \left(\frac{q_f-q_i}{T}\right)^2 - 2 \beta m^2 \left(\frac{q_{f}-q_{0}}{T}\right)^2\right]}~.
\label{fpir} 
\end{eqnarray}
\noindent The above action reduces to that in \cite{sb} in the $\alpha \to 0$ limit. Our next step is to evaluate the constant $\tilde{F}(T, \alpha, \beta)$ which contains the quantum fluctuations. We now use the following identity
\begin{eqnarray}
\langle q_{f}, t_{f}|p\rangle= \int_{-\infty}^{+\infty}dq_0\left<q_{f},t_f|q_{0},t_{0}\right>\left<q_{0},t_{0}|p\right>
\label{ol}
\end{eqnarray}
and the overlaps
\begin{eqnarray}
\langle q_{0},0|p\rangle=\frac{1}{\sqrt{2 \pi \hbar}} exp({\frac{i}{\hbar}pq_0})~;~\langle q_f,T|p\rangle=\frac{1}{\sqrt{2\pi \hbar}}exp\left\{\frac{-i T}{\hbar}\left(\frac{p_{0}^{2}}{2m}-\frac{\alpha}{m} p_{0}^{3}+\frac{\gamma}{m} p_{0}^{4}\right)\right\}~exp\left({\frac{i}{\hbar}pq_f}\right)
\label{ovl}
\end{eqnarray}
with $t_0=0$ and $t_f= T$. Eqs.(\ref{fpir}, \ref{ol}) along with eq.(\ref{ovl}) yields the quantum fluctuations to be  
\begin{eqnarray}
\tilde{F}(T,\alpha, \beta)=\sqrt{\frac{m}{2 \pi i \hbar T}}\left[1+ \frac{3 \alpha m (q_{f}-q_{0})}{T}+\frac{3 i \beta \hbar m}{T}-\frac{6i \alpha^2 \hbar m}{T}-6 \left(\frac{\alpha^2}{2}+\beta \right) \frac{ m^2(q_{f}-q_{0})^2}{T^2}+ \frac{45 \alpha^2 m^2 (q_{f}-q_{0})^2}{2T^2}\right].
\label{confree}
\end{eqnarray}
\noindent Hence from eqs.(\ref{fpir}) and (\ref{confree}) the propagation kernel for a free particle in the GUP framework, containing both the linear and quadratic corrections in momentum, can be recast as
\begin{eqnarray}
\langle q_{f},t_{f} |q_{0}, t_{0} \rangle &=& \sqrt{\frac{m}{2 \pi i \hbar T}}\left[1+ \frac{3 \alpha m (q_{f}-q_{0})}{T}+\frac{3 i \beta \hbar m}{T}-\frac{6i \alpha^2 \hbar m}{T}-6 \left(\frac{\alpha^2}{2}+\beta \right)\frac{ m^2(q_{f}-q_{0})^2}{T^2} \right. \nonumber\\ && \left.+ \frac{45 \alpha^2 m^2 (q_{f}-q_{0})^2}{2T^2}\right]  ~\times e^{\frac{m}{2 T} (q_{f}-q_{0})^2 \left[ 1+ 2 \alpha m \left(\frac{q_f-q_i}{T}\right)+ 8 \alpha^2 m^2 \left(\frac{q_f-q_i}{T}\right)^2 - 2 \beta m^2 \left(\frac{q_{f}-q_{0}}{T}\right)^2\right]}~.
\label{full}
\end{eqnarray}
\noindent 
\noindent Now we proceed to investigate the harmonic oscillator potential with $V(q)= \frac{1}{2} m \omega^2 q^2$ in eq.(\ref{confiaction}). This yields
\begin{eqnarray}
S=\int_0^{T} dt \left[\frac{m}{2} \dot q ^2\left(1+2 \alpha m \dot q + 8 \alpha^2 m^2 \dot q^2 - 2 \beta m^2 \dot q^2 \right)- \frac{1}{2} m \omega^2 q^2 \right]~.
\label{}
\end{eqnarray}
\noindent From the above action one can easily find out the classical equation of motion which reads
\begin{eqnarray}
\ddot q(t)+ 6 \alpha m \dot q(t) \ddot q(t) + 48 \alpha^2 m^2 \dot q^2(t) \ddot q(t) -12 \beta m^2 \dot q^2(t) \ddot q(t) + \omega^2 q(t) =0~.
\label{clseqho}
\end{eqnarray} 
\noindent By solving the above equation we get the classical trajectory of the harmonic oscillator in the presence of the GUP up to order $\mathcal{O}(\alpha^2, \beta)$. The solution can be recast as 
\begin{eqnarray}
q(t)= q_{(0)}(t)+\alpha q_{(1)}(t) + \alpha^2 q_{(2)}(t)+ \beta q_{(3)}(t)
\label{clsoho}
\end{eqnarray}
where 
\begin{eqnarray}
q_{(0)}(t) &=& A \cos(\omega t)+  B \sin(\omega t)~, \nonumber\\  
q_{(1)}(t) &=& C_1 \cos(\omega t)+ C_2 \sin(\omega t) + m \omega (A^2-B^2) \sin(2 \omega t)-2 A B m \omega \cos(2 \omega t)~, \nonumber\\
 q_{(2)}(t) &=& C_3 \cos(\omega t) + C_4 \sin(\omega t) + 3 m^2 \omega^2 (A^2+B^2)(A-Bt\omega) \cos(\omega t)-2 m \omega (BC_1+AC_2)\cos(2\omega t ) \nonumber\\ && - \frac{3}{4} m^2 \omega^2 A(A^2-3 B^2)\cos(3 \omega t)+ 3 m^2 \omega^3 t A (A^2+B^2) \sin(\omega t)+2 m \omega(AC_1-BC_2) \sin(2 \omega t)\nonumber\\ && +\frac{3}{4} m^2 \omega^2 B (B^2-3 A^2) \sin(3 \omega t)~, \nonumber\\  q_{(3)}(t) &=& C_{5} \cos(\omega t)+ C_{6} \sin(\omega t)+ \frac{1}{8} \left[6 m^2 \omega^2 \left(A^2+B^2\right) \left(2 B t \omega-A\right)\cos(\omega t)-3 m^2 \omega^2A (A^2-3B^2)\cos(3 \omega t) \right.  \nonumber\\ && \left. -6m^2 \omega^2 (A^2+B^2)(B+2 A t \omega)\sin(\omega t)+ 3 m^2 \omega^2 B (B^2-3 A^2)\sin(3 \omega t) \right]~.
 \label{solu}
\end{eqnarray}
\noindent The constants $A$, $B$, $C_1$, $C_2$, $C_3$, $C_4$, $C_5$ and $C_6$ read
\begin{eqnarray}
A&=&q_0 \nonumber\\ 
B&=&[q_f-q_0 \cos ( \omega T)]\csc (\omega T)\nonumber\\
C_1&=& 2 m \omega A B \nonumber\\
C_2&=&\frac{1}{\sin(\omega T)} \left[2 AB m \omega \cos(2\omega T)-m \omega (A^2-B^2)\sin(2\omega T)- 2 m \omega A B \cos(\omega T)\right] \nonumber\\ 
C_3&=& -\frac{5}{4}m^2 \omega^2 A B^2+ 2 m \omega A C_2-\frac{9}{4} m^2 \omega^2 A^3 \nonumber\\
C_4&=& -\frac{1}{\sin(\omega T)}\left[C_3 \cos(\omega T)+3 m^2 \omega^2 (A^2+B^2)(A-B\omega T)\cos(\omega T)-2 m \omega (BC_1+AC_2)\cos(2 \omega T)\right. \nonumber\\ && \left.- \frac{3}{4} m^2 \omega^2 A(A^2-3 B^2)\cos(3 \omega T)+ 3 m^2 \omega^3 T A (A^2+B^2)\sin(\omega T)+ 2 m \omega (AC_1-BC_2)\sin(2\omega T)\right. \nonumber\\ && \left.+\frac{3}{4}m^2 \omega^2 B(B^2-3A^2)\sin(3 \omega T)\right] \nonumber\\
C_5&=&\frac{3}{8} m^2 \omega^2 \left(3 A^3- A B^2\right) \nonumber\\ 
C_6&=& -\frac{1}{8 \sin(\omega T)}\left[\left\{6 m^2 \omega^2 (A^2+B^2)(2 B \omega T-A) \cos(\omega T)- 3 n\ m^2 \omega^2 (A^3-3 A^2 B)\cos(3 \omega T) \right. \right. \nonumber\\ && \left. \left. -6 m^2 \omega^2 (A^2 +B^2)(B+2 A \omega t) \sin(\omega T)+ 3 m^2 \omega^2 (B^3-3A^2B)\sin(3 \omega T)\right\}-C_5 \cos(\omega t) \right]~.
\label{cons}
\end{eqnarray}
\noindent Therefore, the classical action for the harmonic oscillator in the framework of the GUP algebra (\ref{gupu}) can be obtained by using eqs.(\ref{clsoho}), (\ref{solu}) and (\ref{cons}) in eq.(\ref{clseqho}). This yields
\begin{eqnarray}
S_{c}= S_{c}(0)+S_{c}(\alpha)+S_{c}(\alpha^2) +S_{c}(\beta)~.
\label{hofa}
\end{eqnarray}

\noindent Here $ S_{c}(0)$ is the classical action for the ordinary harmonic oscillator. $S_{c}(\alpha)$, $S_{c}(\alpha^2)$ and $S_{c}(\beta)$ are the corrections due to the presence of the GUP. The forms of $ S_{c}(0)$, $S_{c}(\alpha)$, $S_{c}(\alpha^2)$ and $S_{c}(\beta)$ are

\begin{eqnarray}
S_c(0)=\frac{1}{2} m w \csc (T w) \left[\left(q_0^2+q_f^2\right) \cos (T w)-2 q_0 q_f\right]
\label{clacoho}
\end{eqnarray}
\begin{eqnarray}
S_c(\alpha)=- \frac{ \alpha}{6} m^2 w^2 (q_0-q_f) \csc ^2(T w) \left[\left(q_0^2+q_0 q_f+q_f^2\right) \cos (2 T w)-12 q_0 q_f \cos (T w)-q_0 q_f+5 (q_0^2+q_f^2)\right]
\label{clacho1}
\end{eqnarray}
\begin{eqnarray}
S_c(\alpha^2) &=& \frac{ \alpha^2 }{16} m^3 w^3 \csc ^4(T w) \left[\left(q_0^4+q_f^4\right) \sin (4 T w)-4 q_0 q_f \left(21 q_0^2-20 q_0 q_f+21 q_f^2\right) \sin (T w) \right. \nonumber\\ && \left. -4 q_0 q_f \left(5 q_0^2-4 q_0 q_f+5 q_f^2\right) \sin (3 T w) +24 q_0^2 q_f^2 T w \cos (2 T w)-48 q_0 q_f T w \left(q_0^2+q_f^2\right) \cos (T w) \right.\nonumber\\ && \left. +12 T w \left(q_0^4+4 q_0^2 q_f^2+q_f^4\right)  +4 \left(6 q_0^4-8 q_0^3 q_f+23 q_0^2 q_f^2-8 q_0 q_f^3+6 q_f^4\right) \sin (2 T
   w)\right]
\label{clacho2}
\end{eqnarray}
\begin{eqnarray}
S_{(\beta)} &=& - \frac{\beta}{32} m^3 w^3 \csc ^4(T w) \left[\left(q_0^4+q_f^4\right) \sin (4 T w)-44 q_0 q_f \left(q_0^2+q_f^2\right) \sin (T w)-12 q_0 q_f \left(q_0^2+q_f^2\right) \sin (3 T w) \right. \nonumber\\ && \left.+24 q_0^2 q_f^2 T w \cos (2 T w)  -48 q_0 q_f T w \left(q_0^2+q_f^2\right) \cos (T w)+12 T w \left(q_0^4+4 q_0^2 q_f^2+q_f^4\right) \right. \nonumber\\ && \left. +4 \left(2 q_0^4+15 q_0^2 q_f^2+2 q_f^4\right) \sin (2 T w)\right]~.
\label{clacho3}
\end{eqnarray}
\noindent It is reassuring to note that we recover the free particle classical action (\ref{clssaction}) in the limit $\omega \to 0$. Therefore, the propagator for the harmonic oscillator reads
\begin{eqnarray}
\langle q_{f},t_{f}|q_{0},t_{0}\rangle= \sqrt{\frac{m \omega}{2 \pi i \hbar \sin(\omega T)}}\tilde{F_1}(T, \alpha, \beta) ~ e^{\frac{i}{\hbar} S_{c}}~.
\label{fprho}
\end{eqnarray}
\noindent Now we will calculate the quantum fluctuations $\tilde{F_1}$ from the Schr\"{o}dinger equation in the subsequent section . 
\section{Calculation of the quantum fluctuation}

\noindent In this section we apply a different approach to evaluate the explicit form of the kernel of a particle moving in a harmonic potential in the GUP framework. We calculate the Feynman propagator and the quantum fluctuations $\tilde{F_1}$ from the Schr\"{o}dinger equation upto order $\mathcal{O}(\alpha, \beta)$. Note that in this section though we give the complete expression of eigenfunctions and energy eigenvalues retaining the terms in $\alpha^2$ in the the final expression of the quantum fluctuation for harmonic oscillator we neglect the terms of the order $\mathcal{O}(\alpha^2, \beta^2)$.

\noindent To do this first we write the Schr\"{o}dinger equation for the harmonic oscillator bearing the GUP effects for both linear and quadratic corrections in momentum. This reads
\begin{equation}
\left[ -\frac{\hbar^2}{2 m} \frac{\partial^2}{\partial q^2} -\frac{i \alpha \hbar^3}{m} \frac{\partial^3}{\partial q^3} + \frac{\alpha^2 \hbar^4}{2m} \frac{\partial^4}{\partial q^4} + \frac{\beta \hbar^4}{m} \frac{\partial^4}{\partial q^4}  + \frac{1}{2} m \omega^2 q^2 + {\cal O} (\beta^2)\right] \psi_{n} (q) = E_n \psi_n (q)
\label{k-schrodinger}
\end{equation}
where $\psi_n (q)$ and $E_n$ are $n^{th}$ order eigenfunction and eigenvalue of the Schr\"{o}dinger equation. Hence the Feynman propagator $\langle q_{f},t_{f}|q_{0},t_{0}\rangle $ can be recast as
\begin{eqnarray}
\langle q_{f},t_{f}|q_{0},t_{0}\rangle = \sum_{n} \psi_n (q_f) \psi_n^* (q_0) e^{-(i / \hbar) E_n (t_f - t_0)}~.
\label{fynpro}
\end{eqnarray}

\noindent We now solve the Schr\"{o}dinger equation (\ref{k-schrodinger}) by treating the GUP contributions as time independent perturbations. Then the perturbation piece of the Hamiltonian upto the order $\mathcal{O}(\alpha^2, \beta)$ can be written as 
\begin{eqnarray}
H(\alpha, \alpha^2, \beta)=-\frac{\alpha}{m} p_{0}^{3}+\frac{1}{m}\left(\frac{\alpha^2}{2}+\beta\right) p_{0}^{4}+\frac{1}{2}m \omega^2 q^2  +\mathcal O( \alpha \beta, \alpha^3, \beta^2)~.
\label{pethamil}
\end{eqnarray}
We can now obtain the eigenstates and eigenvalues by applying time independent perturbation theory. This yields
\begin{eqnarray}
	\label{ks}
\psi_n(q) &=& \phi_n(q) -\frac{i \alpha}{m \hbar \omega} \left(\frac{\hbar m \omega}{2} \right)^{\frac{3}{2}} \left[\frac{\sqrt{n(n-1)(n-2)}}{3} \phi_{n-3}(q) -3 n \sqrt{n} \phi_{n-1}(q)-3 (n+1)\sqrt{n+1} \phi_{n+1}(q) \right. \nonumber\\ && \left. +\frac{(n+1)(n+2)(n+3)}{3}\phi_{n+3}(q) \right] +\left(\frac{\alpha^2}{2}+\beta \right)(m \hbar \omega)    \left[ \frac{(2 n + 3) \sqrt{(n+1) (n + 2)}}{4} \phi_{n + 2} (q) \right.  \nonumber\\ && \left. - \frac{(2 n - 1) \sqrt{n (n - 1)}}{4} \phi_{n - 2} (q) +  \frac{\sqrt{n (n - 1) (n - 2) (n - 3)}}{16} \phi_{n-4} (q) - \frac{\sqrt{ (n + 1) (n + 2) (n + 3) (n + 4)}}{16} \phi_{n+4} (q)    \right] \nonumber\\ && +{\cal O} (\alpha^3) +{\cal O} (\beta^2) + {\cal O} (\alpha \beta)    \\    \nonumber \text{and} \nonumber\\
&& E_n = \left(n + \frac{1}{2} \right) \hbar \omega \left[ 1 + \frac{3 (2 n^2 + 2 n + 1)}{2 (2 n + 1)} \left(\frac{\alpha^2}{2}+\beta\right)( m \hbar \omega) \right] + {\cal O} (\beta^2),
\label{k-schrodinger-3}
\end{eqnarray}
where $n = 0, 1, 2, \cdots$ and 
\begin{equation}
\phi_n (q) = \frac{1}{\sqrt{2^n n!}}  \left( \frac{m \omega}{\pi \hbar} \right)^{1/4} H_n \left( \sqrt{\frac{m \omega}{\hbar}} q \right) \exp \left[ - \frac{m \omega}{2 \hbar} q^2 \right].
\label{k-schrodinger-4}
\end{equation}
\noindent These are the complete form of the eigenstates and energy eigenvalues of harmonic oscillator in the presence of the GUP, with the general GUP structure containing both linear and quadratic contributions in momentum upto order $\mathcal{O}(\alpha, \alpha^2, \beta)$. Using eqs.(\ref{ks}) and (\ref{k-schrodinger-3})  in eq.(\ref{fynpro}), the Feynman propagator reads 
\begin{eqnarray}
\langle q_{f},t_{f}|q_{0},t_{0}\rangle &=& J + \frac{i \alpha}{m \omega \hbar} \left(\frac{m \omega \hbar}{2}\right)^{\frac{3}{2}} \left[M_1+M_2\right] + (\beta m \hbar \omega) \left[N_1 + N_2 \right] + {\cal O} (\beta^2) +\mathcal{O}(\alpha^2, \beta)
\label{propa}
\end{eqnarray}

\noindent where
\begin{eqnarray}
J &=& \sum_{n=0}^{\infty} \phi_n (q_f) \phi_n (q_0) \exp \left[ -\frac{i}{\hbar} \left( n + \frac{1}{2} \right) \hbar \omega T \left\{ 1 + \frac{3 (2 n^2 + 2 n + 1)}{2 (2 n + 1)} (\beta m \hbar \omega) \right\} \right]    \nonumber\\
M_1 &=& \left[ \sum_{n=3}^{\infty} \frac{\sqrt{n (n-1) (n-2)}}{3}\left[\phi_n(q_f) \phi_{n-3}(q_0)-\phi_n(q_0) \phi_{n-3}(q_f)\right] \right. \nonumber\\ && \left. +  \sum_{n=0}^{\infty} \frac{\sqrt{(n+1)(n+2)(n+3)}}{3} \left[\phi_{n+3}(q_0) \phi_n(q_f)- \phi_{n+3}(q_f)\phi_n(q_0)\right] \right] \times \exp{\left[-\frac{i}{\hbar}\left(n+\frac{1}{2}\right) \hbar \omega T\right]} \nonumber\\ M_2 &=& - 3 \left[\sum_{n=1}^{\infty} n \sqrt{n} \left[\phi_{n}(q_f) \phi_{n-1}(q_0)-\phi_{n-1}(q_f)\phi_n(q_0) \right]+ \sum_{n=0}^{\infty} (n+1)\sqrt{n+1} \left[\phi_n(q_f) \phi_{n+1}(q_0)- \phi_{n+1}(q_f) \phi_n(q_0)\right]\right] \nonumber\\ && \times \exp{\left[-\frac{i}{\hbar}\left(n+\frac{1}{2}\right) \hbar \omega T\right]} \nonumber\\  N_1 &=& \Bigg[\sum_{n=0}^{\infty} \frac{(2 n + 3) \sqrt{(n + 1) (n + 2)}}{4} \left[ \phi_n (q_f) \phi_{n + 2} (q_0) + \phi_n (q_0) \phi_{n+2} (q_f) \right]  \nonumber \\ &&  - \sum_{n=2}^{\infty} \frac{(2 n - 1) \sqrt{n  (n -1)}}{4} \left[ \phi_n (q_f) \phi_{n - 2} (q_0) + \phi_n (q_0) \phi_{n-2} (q_f) \right] \Bigg] \exp \left[-\frac{i}{\hbar} \left(n + \frac{1}{2} \right) \hbar \omega T \right]   \nonumber \\  N_2 &=& \Bigg[\sum_{n=4}^{\infty} \frac{\sqrt{n(n-1) (n - 2) (n - 3)}}{16} \left[ \phi_n (q_f) \phi_{n - 4} (q_0) + \phi_n (q_0) \phi_{n - 4} (q_f)\right]  \nonumber \\ &&  - \sum_{n=0}^{\infty} \frac{ \sqrt{(n +1) (n + 2) (n + 3) (n + 4)}}{16} \left[ \phi_n (q_f) \phi_{n + 4} (q_0) + \phi_n (q_0) \phi_{n+4} (q_f) \right] \Bigg]. \nonumber \\ &&   \times \exp \left[-\frac{i}{\hbar} \left(n + \frac{1}{2} \right) \hbar \omega T \right]~.
\label{con}
\end{eqnarray}
\noindent Now using the exact form of $\phi_{n}(q)$ given by the eq.(\ref{k-schrodinger-4}) in eq. (\ref{con}), we have
\begin{eqnarray}
J &=& \sqrt{\frac{m \omega}{\pi \hbar}} \exp{\left[-\frac{m \omega}{2 \hbar}(q_0^2+q_f^2)\right]}	
\exp{\left(-\frac{i \omega T}{2}\right)}\sum_{n=0}^{\infty} \left(\frac{\exp{(-i \omega T)}}{2}\right)^n \frac{1}{n!}	H_n \left( \sqrt{\frac{m \omega}{\hbar}} q_0 \right) H_n \left( \sqrt{\frac{m \omega}{\hbar}} q_f \right) \nonumber \\ && \left[1- \frac{3 i \beta m \omega^2 \hbar T}{4} (2n^2+2n+1)\right] 
\nonumber \\
M_1 &=& -\sqrt{\frac{m \omega}{\pi \hbar}} \frac{i}{3\sqrt{2}} \exp{\left[-\frac{m \omega}{2 \hbar}(q_0^2+q_f^2)\right]} \exp{(-2 i \omega T)} \sin\left(\frac{3 \omega T}{2}\right)\sum_{n=0}^{\infty}\left(\frac{\exp{(-i \omega T)}}{2}\right)^n \frac{1}{n!} \nonumber\\ && \left[H_{n+3} \left( \sqrt{\frac{m \omega}{\hbar}} q_f \right)H_n \left( \sqrt{\frac{m \omega}{\hbar}} q_0 \right)-H_{n+3} \left( \sqrt{\frac{m \omega}{\hbar}} q_0 \right)H_n \left( \sqrt{\frac{m \omega}{\hbar}} q_f \right)\right]
\nonumber \\
M_2 &=& \sqrt{\frac{m \omega}{\pi \hbar}}  3 i \sqrt{2} \exp{\left[-\frac{m \omega}{2 \hbar}(q_0^2+q_f^2)\right]} \exp{(- i \omega T)} \sin\left(\frac{ \omega T}{2}\right)\sum_{n=0}^{\infty}(n+1)\left(\frac{\exp{(-i \omega T)}}{2}\right)^n \frac{1}{n!} \nonumber\\ && \left[H_{n+1} \left( \sqrt{\frac{m \omega}{\hbar}} q_f \right)H_n \left( \sqrt{\frac{m \omega}{\hbar}} q_0 \right)-H_{n+1} \left( \sqrt{\frac{m \omega}{\hbar}} q_0 \right)H_n \left( \sqrt{\frac{m \omega}{\hbar}} q_f \right)\right]	
\nonumber \\
N_1 &=& \sqrt{\frac{m \omega}{\pi \hbar}}  3 i \sqrt{2} \exp{\left[-\frac{m \omega}{2 \hbar}(q_0^2+q_f^2)\right]} exp{(-\frac{3i \omega T}{2})} \sin(\omega T) \sum_{n=0}^{\infty}\left(\frac{\exp{(-i \omega T)}}{2}\right)^n \frac{1}{n!} \nonumber\\ && \left[H_{n+2} \left( \sqrt{\frac{m \omega}{\hbar}} q_f \right)H_n \left( \sqrt{\frac{m \omega}{\hbar}} q_0 \right)+H_{n+2} \left( \sqrt{\frac{m \omega}{\hbar}} q_0 \right)H_n \left( \sqrt{\frac{m \omega}{\hbar}} q_f \right)\right]~.
\label{con2}
\end{eqnarray} 
\noindent Similarly $N_2$ can be recast in terms of the Hermite polynomials.

\noindent Now to evaluate the constants $J$, $M_1$, $M_2$,$N_1$ and $N_2$, we use the extended Mehler's formula \cite{app}
\begin{eqnarray}
\label{mehler-1}
&&\sum_{k=0}^{\infty} \frac{t^k}{k!} H_{k+m} (x) H_{k+n} (y) = (1 - 4 t^2)^{-(m + n + 1) / 2} \exp \left[ \frac{4 t x y - 4 t^2 (x^2 + y^2)}{1 - 4 t^2} \right]          \\    \nonumber
&& \hspace{1.5cm}  \times \sum_{k = 0}^{\min (m,n)} 2^{2 k} k! \left(   \begin{array}{c} m  \\  k   \end{array}   \right)    \left(   \begin{array}{c} n  \\  k   \end{array}   \right)   t^k 
H_{m - k} \left( \frac{x - 2 t y}{\sqrt{1 - 4 t^2}} \right)    H_{n - k} \left( \frac{y - 2 t x}{\sqrt{1 - 4 t^2}} \right)~.
\end{eqnarray}
\noindent Using this we get
\begin{eqnarray}
\label{k-schrodinger-9}
&& J =  \sqrt{\frac{m \omega}{2 \pi i \hbar \sin \omega T}} e^{\frac{i}{\hbar} S_0} \tilde{J}~~,
M_1 =  \sqrt{\frac{m \omega}{2 \pi i \hbar \sin \omega T}} e^{\frac{i}{\hbar} S_0} \tilde{M}_1~~, M_2 =  \sqrt{\frac{m \omega}{2 \pi i \hbar \sin \omega T}} e^{\frac{i}{\hbar} S_0} \tilde{M}_2~~,\nonumber \\ &&
N_1 =  \sqrt{\frac{m \omega}{2 \pi i \hbar \sin \omega T}} e^{\frac{i}{\hbar} S_0} \tilde{N}_1~~, N_2 =  \sqrt{\frac{m \omega}{2 \pi i \hbar \sin \omega T}} e^{\frac{i}{\hbar} S_0} \tilde{N}_2~~, \hspace{1.0cm}
\end{eqnarray}

where 
\begin{eqnarray}
	\label{j}
	&&\tilde{J} = 1 - \frac{3 i \beta m \omega^2 T}{8 \hbar \sin^4 \omega T} \Bigg[ -3 i \hbar m \omega (q_0^2 + q_f^2) \sin 2 \omega T + m^2 \omega^2 (q_0^2 + q_f^2 -2 q_0 q_f \cos \omega T)^2      \nonumber    \\
	&& \hspace{2.0cm}       + 4 i \hbar m \omega \sin \omega T \left( 2 + \cos 2 \omega T \right) q_0 q_f - \hbar^2 \sin^2 \omega T \left( 2 + \cos 2 \omega T \right)  \Bigg]                        
	\end{eqnarray}
\begin{eqnarray}
\label{M1}
\tilde{M}_1 &=&   \frac{1}{3} \sqrt{\frac{m \omega}{2 \hbar}} \frac{\sin{\frac{3 \omega T}{2}}}{\hbar \sin^2{\omega T} \sin{\frac{\omega T}{2}}} (q_0-q_f) \left[-m \omega \left(q_0^2+4 q_0 q_f +q_f^2\right)+ 2 m \omega \left(q_0^2+ q_0 q_f +q_f^2\right) \cos{\omega T} \right. \nonumber\\ && \left.  - 3 i \hbar \sin{\omega T} \right] \nonumber\\  
\tilde{M}_2 &=& -\frac{3\sqrt{2}}{8 \hbar} \sqrt{\frac{m \omega}{\hbar}} \frac{(q_0-q_f)}{\sin^2{\frac{\omega T}{2} \cos^2{\frac{\omega T}{2}}}}\left[-i \hbar \sin{2 \omega T}+ m \omega (q_0^2-2q_0 q_f \cos{\omega T}+ q_f^2)-i \hbar \sin{\omega T}\right]
\end{eqnarray}
\begin{eqnarray}
\label{N1}
\tilde{N}_1 &=& - \frac{i}{8 \hbar^2 \sin^3 \omega T} \left[ -4 m^2 \omega^2 q_0 q_f (q_0^2 + q_f^2) (3 + \cos  2 \omega T) + 3 \hbar^2 (\cos 3 \omega T - \cos \omega T)  \right. \nonumber  \\ && \left. + 4 m \omega \cos \omega T \left\{ m \omega (q_0^4 + 6 q_0^2 q_f^2 + q_f^4) + 12 i \hbar q_0 q_f \sin \omega T \right\} - 3 i \hbar m \omega (q_0^2 + q_f^2) (5 \sin \omega T + \sin 3 \omega T)   \right]
\nonumber \\
\tilde{N}_2 &=& - \frac{i \cos \omega T}{16 \hbar^2 \sin^3 \omega T} \left[ 12 m^2 \omega^2 q_0^2 q_f^2 - 3 \hbar^2 (1 - \cos 2 \omega T) + 2 m \omega \left\{ m \omega \cos 2 \omega T (q_0^4 + q_f^4) \right. \right.  \nonumber \\ && \left.  \left.   - 4 m \omega  q_0 q_f (q_0^2 + q_f^2) \cos \omega T - 6 i \hbar \sin \omega T \left\{ (q_0^2 + q_f^2) \cos \omega T - 2 q_0 q_f \right\} \right\}    \right].
\end{eqnarray} 


Therefore, $\langle q_{f},t_{f}|q_{0},t_{0}\rangle$ can be recast as (upto $\mathcal{O}(\alpha, \beta)$)
\begin{eqnarray} 
\langle q_f, t_f | q_0, t_0 \rangle &=& \sqrt{\frac{m \omega }{2 \pi i \hbar \sin{\omega T}}} \left[1+ \alpha f(q_0, q_f; T)+ \beta ~g(q_0, q_f; T)+ \mathcal{O}(\alpha, \beta) \right] e^{\frac{i}{\hbar}(S_c(0)+S_c(\alpha)+S_c(\beta))}
\label{}
\end{eqnarray}
where
\begin{eqnarray}
S_c(0) &=& \frac{m \omega }{2 } \csc{\omega T} \left[(q_0^2+q_f^2) \cos{\omega T}- 2 q_0 q_f \right]  \nonumber\\
S_c(\alpha) &=& -\frac{\alpha}{6} m^2 \omega^2 (q_0-q_f) \csc^2{\omega T} \left[(q_0^2+ q_0 q_f+ q_f^2) \cos{2 \omega T} -12 q_0 q_f \cos{\omega T}-q_0 q_f + 5 (q_0^2+q_f^2)\right] \nonumber
\label{casm}
\end{eqnarray}
\begin{eqnarray}
S_c(\beta) &=& - \frac{\beta m^3 \omega^3}{32 } \csc^4 \omega T \Bigg[ \left\{ 12 \omega T + 8 \sin 2 \omega T + \sin 4 \omega T \right\} (q_0^4 + q_f^4)               \\     \nonumber
&& \hspace{1.5cm} -4 \left\{ 12 \omega T \cos \omega T + 11 \sin \omega T + 3 \sin 3 \omega T \right\} q_0 q_f (q_0^2 + q_f^2)                                           \\     \nonumber
&& \hspace{4.5cm}   + 12 \left\{ 4 \omega T + 2 \omega T \cos 2\omega T + 5 \sin 2 \omega T \right\} q_0^2 q_f^2    \Bigg]
\label{action}
\end{eqnarray}
with the functions f and g being given by
\begin{eqnarray}
&& f(q_0, q_f; T) = - (q_0-q_f)m \omega \csc^2{\omega T}\left[\sin{\omega T}+\sin{2 \omega T}\right] \nonumber\\
&& g(q_0, q_f: T) = \frac{3 i \hbar m \omega}{8 \sin^2 \omega T}  \left( 2 \omega T + 5 \sin \omega T \cos \omega T + \omega T \cos 2 \omega T \right)                   \\     \nonumber
&& \hspace{2.8cm} - \frac{3 m^2 \omega^2}{8 \sin^3 \omega T} \Bigg[ 2 \omega T \left\{ 3 \cos \omega T (q_0^2 + q_f^2) -  2 (2 + \cos 2 \omega T )q_0 q_f  \right\}                     \\     \nonumber
&& \hspace{5.0cm} + 10 \sin \omega T (q_0^2 + q_f^2 - 2 q_0 q_f \cos \omega T) - 6 \sin^3 \omega T (q_0^2 + q_f^2)       \Bigg].
\end{eqnarray}
\noindent  Note that in this method we calculate the exact expression for the quantum fluctuation up to first order in $\alpha, \beta$. This calculation can be extended for higher order in $\alpha^2$.


\section{Conclusion}
We now summarize the results in this paper. In this paper we have constructed the path integral formalism of the propagation kernel in presence of the generalized uncertainty principle incorporating both the contributions proportional to linear and quadratic terms in momentum. We obtained the action of a nonrelativistic particle moving in an arbitrary potential in the framework of the generalized uncertainty principle. After getting the general form of the action we have moved on to investigate the free particle and harmonic oscillator systems. From the free particle analysis, we have seen that the action imposes an upper bound on the free particle velocity which depends on the mass of the particle. This feature is consistent with the results obtained in earlier \cite{sp,sb,fgsg}. Moreover the fact that the particle velocity must be real and finite leads us to a relation between the parameters $\alpha$ and $\beta$. We show that $\beta> 4\alpha^2$. This is an interesting result in our paper. Then we have calculated the Feynman propagator for a harmonic oscillator. In the limiting case $\omega \to 0$, the classical action for the  harmonic oscillator reduces to the free particle result. We have explored another approach to get the propagation kernel. We have constructed the Schr\"odinger equation for a harmonic oscillator in the framework of the generalized uncertainty principle. Solving the Schr\"odinger equation we have got expressions for $n$-th order eigenfunction and energy eigenvalue bearing the effects of the generalized uncertainty principle. Using these results we derive the expression for the propagation kernel for the harmonic oscillator. We have obtained the explicit form of the quantum fluctuations upto first order in $\alpha$ and $\beta$. These results would be important to derive the thermodynamics of the harmonic oscillator system in the general uncertainty principle framework. This we hope to report in future.    

\section{Appendix-I}
\noindent The most general algebra \cite{sd} for the commutation relation between position $q_j$ and its conjugate momentum $p_j$ with linear and quadratic modifications in momentum reads
\begin{eqnarray}
[q_i, p_j] &=& i \hbar \left(\delta_{ij}+\delta_{ij} \alpha_1 p+\alpha_2 \frac{p_i p_j}{p} +\beta_1 \delta_{ij} p^2 +\beta_2 p_i p_j \right)~.
\label{GUPRR}
\end{eqnarray} 
\noindent Therefore the coordinates and its conjugate momentum follow the Jacobi identity
\begin{eqnarray}
-[[q_i,q_j],p_k]=[[q_j, p_k], q_i]+[[p_k,q_i], q_j]=0~.
\label{jaci}
\end{eqnarray}
Now we expand the right hand side of the Jacobi identity and using eq.(\ref{GUPRR}), we get 
\begin{eqnarray}
i \hbar\left\{-\alpha_1 \delta_{jk} [q_i,p]- \alpha_2 \left([q_i,p_j]p_k p^{-1}+ p_j [q_i, p_k]p^{-1}+p_jp_k[q_i,p^{-1}]\right)- \beta_1 \delta_{jk}\left([q_i, p_l]p_l+p_l[q_i,p_l]\right)\right. \nonumber\\ \left.- \beta_2\left([q_i,p_j]p_k+p_j[q_i,p_k]\right)\right\}-\left(i\leftrightarrow j\right)=0~.
\label{jacid}
\end{eqnarray}
We can easily evaluate the following commutator upto $\mathcal{O}(p)$,
\begin{eqnarray}
[q_i,p] &=& i\hbar \left\{p_ip^{-1}+(\alpha_1+\alpha_2)p_i\right\}
\label{comre1}
\end{eqnarray}
and,
\begin{eqnarray}
[q_i,p^{-1}] &=& -i \hbar p_ip^{-3} \left\{1+(\alpha_1+\alpha_2)p\right\}~.
\label{comre2}
\end{eqnarray}
\noindent Using the above commutation relations in eq.(\ref{jacid}), we get
\begin{equation}
\left\{(\alpha_1-\alpha_2) p^{-1}+(\alpha_1^2+2 \beta_1-\beta_2)\right\} \left(p_i \delta_{jk}- p_j \delta{ik}\right)=0~.
\label{comre3}
\end{equation}
\noindent Thus the above equation is satisfied only when $\alpha_1=\alpha_2= \alpha$ ($\alpha > 0$ \cite{jm} )
and $\beta_2=2\beta_1+\alpha_1^2$. Now from dimensional analysis we have $\beta_1 \sim \alpha^2$. Let $\beta_1=n \alpha^2$, where $n$ is positive number. Then we have $\beta_2=(2n+1) \alpha^2$. Note that in \cite{sd} $\beta_1=\alpha^2$ (that is $n=1$) has been taken into account for mathematical simplicity. Putting the values of $\beta_1$ and $\beta_2$ in eq.(\ref{GUPRR}), the commutation relation takes the form as
\begin{eqnarray}
[q_i, p_j] &=& i \hbar\left[\delta_{ij}-\alpha \left(p \delta_{ij}+ \frac{p_ip_j}{p}\right)+ n \alpha^2 p^2 \delta_{ij}+ \left(2 n +1\right) \alpha^2 p_ip_j\right]~.
\label{comre4}
\end{eqnarray}
\noindent Now the most general form of the momentum $p_j$ in terms of $p_{0j}$ can be written as 
\begin{eqnarray}
p_j &=& p_{0j} + a p_0 p_{0j}+ b p_0^2 p_{0j}~,
\label{prel} 
\end{eqnarray}
\noindent where $a \sim \alpha$ and $b \sim \alpha^2$. Hence the commutation relation can be recast as
\begin{eqnarray}
[q_i, p_j] &=& [q_i, p_{0j}+ a p_0 p_{0j}+ b p_0^2 p_{0j}] \nonumber\\ 
&=& i \hbar \delta_{ij}+i \hbar a \left(p\delta_{ij}+p_i p_j p^{-1}\right) +i \hbar \left(2 b-a^2 \right)p_i p_j +(b-a^2) p^2 \delta_{ij}~.
\label{comrel44}
\end{eqnarray}
\noindent Comparing the above relation with (\ref{comre4}), finally we get $a=-\alpha$, $n \alpha^2= b -a^2$ and $(2n+1)\alpha^2= 2b- a^2$. Hence
\begin{eqnarray}
b= (n+1) \alpha^2~.
\label{newrel}
\end{eqnarray} 
\noindent Note that if we take $n=1$ for mathematical simplicity, then we get $\beta = 2 \alpha^2$ \cite{sd}. Now using the above relations we define two parameters, bearing the signature of the GUP as $a=-\alpha$ and $(n+1)\alpha^2=\beta$. Therefore eq.(\ref{prel}) yields
\begin{eqnarray}
p_j &=& p_{0j} -\alpha p_0 p_{0j}+ \beta p_0^2 p_{0j}~,
\end{eqnarray}
where $\beta= (n+1) \alpha^2$. This is eq.(\ref{GUPR}) in this paper. In our analysis, eq.(\ref{rel}) shows that 
\begin{eqnarray}
\beta > 4 \alpha^2 ~.
\label{rel2}
\end{eqnarray}
This implies $n>3$.

\end{document}